\documentclass[10pt,final]{iopart}

\usepackage{graphicx}	% Including figure files
\usepackage{iopams}
\expandafter\let\csname equation*\endcsname\relax
\expandafter\let\csname endequation*\endcsname\relax
\usepackage{amsmath}	% Advanced maths commands
\usepackage{amssymb}	% Extra maths symbols
\usepackage[colorlinks=true,
            linkcolor=blue,
            citecolor=green,
            filecolor=magenta,
            urlcolor=cyan]{hyperref}
\usepackage[numbers,sort&compress]{natbib}
\usepackage{aasmacros}

\newcommand{\pc}{\mathrm{pc}}
\newcommand{\msol}{M_\odot}

\begin{document}

%%%%%%%%%%%%%%%%%%%%%%%%%%%%%%%%%%%%%%%%%%%%%%%%%%

%%%%%%%%%%%%%%%%%%% TITLE PAGE %%%%%%%%%%%%%%%%%%%

% Title of the paper, and the short title which is used in the headers.
% Keep the title short and informative.
\title[Host demographics of SMBHBs]{Host Galaxy Demographics Of Individually Detectable Supermassive Black-hole Binaries with Pulsar Timing Arrays}

% The list of authors, and the short list which is used in the headers.
% If you need two or more lines of authors, add an extra line using \newauthor
\author{Katharine Cella$^1$, Stephen~R.~Taylor$^1$ and Luke~Zoltan~Kelley$^2$,$^3$}
\address{$^1$ Department of Physics \& Astronomy, Vanderbilt University,
2301 Vanderbilt Place, Nashville, TN 37235, USA}
\ead{katieanncella@gmail.com}
\ead{Corresponding author, stephen.r.taylor@vanderbilt.edu}
\address{$^2$ Department of Astronomy, University of California at Berkeley, Berkeley, CA 94720}
\address{$^3$ Center for Interdisciplinary Exploration and Research in Astrophysics (CIERA), Northwestern University, Evanston, IL 60208, USA}
\ead{lzkelley@berkeley.edu}

% These dates will be filled out by the publisher
%\date{Accepted XXX. Received YYY; in original form ZZZ}

% Enter the current year, for the copyright statements etc.
%\pubyear{2022}

% Don't change these lines
%\begin{document}
%\label{firstpage}
%\maketitle

\begin{abstract}
Massive black hole binaries (MBHBs) produce gravitational waves (GWs) that are detectable with pulsar timing arrays. We determine the properties of the host galaxies of simulated MBHBs at the time they are producing detectable GW signals. The population of MBHB systems we evaluate is from the \textit{Illustris} cosmological simulations taken in tandem with post processing semi-analytic models of environmental factors in the evolution of binaries. Upon evolving to the GW frequency regime accessible by pulsar timing arrays, we calculate the detection probability of each system using a variety of different values for pulsar noise characteristics in a plausible near-future International Pulsar Timing Array dataset. %We average over multiple realizations of the Universe by re-sampling the host galaxy properties using a kernel density estimator to approximate the statistical distributions of the Universe. Excitingly, 
We find that detectable systems have host galaxies that are clearly distinct from the overall binary population and from most galaxies in general. With conservative noise factors, we find that host stellar metallicity, for example, peaks at $\sim2Z_\odot$ as opposed to the total population of galaxies which peaks at $\sim0.6Z_\odot$. Additionally, the most detectable systems are much brighter in magnitude and more red in color than the overall population, indicating their likely identity as large ellipticals with diminished star formation. These results can be used to develop effective search strategies for identifying host galaxies and electromagnetic counterparts following GW detection by pulsar timing arrays.
\end{abstract}

%\begin{keywords}
%supermassive black holes -- keyword2 -- keyword3
%\end{keywords}

%%%%%%%%%%%%%%%%%%%%%%%%%%%%%%%%%%%%%%%%%%%%%%%%%%

%%%%%%%%%%%%%%%%% BODY OF PAPER %%%%%%%%%%%%%%%%%%

\section{Introduction}

It is now well accepted that massive black holes occupy the centers of most galaxies \citep{1995ARA&A..33..581K}. By undergoing repeated mergers over cosmic time, galaxies grow in a hierarchical paradigm through mergers, as well as through the accretion of dark matter and gas from cosmic web filaments \citep{Press+Schechter-1974, Blumenthal+1984, Cole+2000}. When such galaxies merge, their central black holes initially experience dynamical friction within the resulting galactic merger remnant, slowly sinking to the center of the gravitational potential of the merged galaxy through the ensemble of many long-range scattering events \citep{1980Natur.287..307B, Merritt_2005, Antonini_2011}. At this stage, prior to the two black holes becoming gravitationally bound, they are referred to as a ``dual'' system.

There are a variety of interactions between the two massive black-holes (MBHs) and their ambient environment that can `harden' it, i.e., dissipate orbital energy and lead to a reduction in black-hole separation. Broadly speaking, these mechanisms are stellar scattering, torques from a gaseous circumbinary disk, and the emission of gravitational waves \citep[GWs;][]{1980Natur.287..307B}; and at times three-body interactions from subsequent galaxy mergers \citep{Makino+1994, Blaes+2002, Hoffman+2007, Bonetti+2016}. 
Once the binary separation is sufficiently small such that the mass enclosed is lower than the total mass of the black holes (typically at $\sim 1 - 10$~pc), the dynamics will be driven by the black holes themselves and they become gravitationally-bound as a ``binary''.
From the merging of host galaxies to formation of a binary may take several Gyrs \citep{Kelley_2017a}, and significant uncertainties remain regarding the scattering of stars in the ``final parsec'' \citep{Milosavljevic+Merritt-2003, Berczik+2006, Khan+2011, Vasiliev+2015}.

At separations of $\lesssim 1$~pc, a circumbinary disk can form from accreting gas that is channeled into the central galactic regions as a result of the turbulent conditions following the galaxy merger \citep{1996ApJ...471..115B}. If the black holes are of comparable mass, the torque exerted by the binary can carve out a cavity in the center of the circumbinary disk, leading to a depleted region within which the MBHB is surrounded by a dam of gas \citep{1977ApJ...216..822P}. This dam can leak tendrils of gas onto the black holes, leading to the formation of circum-single disks (or `mini-disks') around each \citep{1994ApJ...421..651A, 2002A&A...387..550G}.
For decades it was believed that the torques exerted between the accretion disks and binary leads to energy and angular momentum extraction from the binary, further contributing to the system's hardening \citep{1979MNRAS.186..799L, Gould_2000}.  More recently, \citet{Miranda+2017} and \citet{Munoz+2019} have shown that in some cases the presence of gas can expand the binary instead of contracting it.  While this result has been confirmed by other groups \citep{Moody+2019, Duffell+2020, Siwek+2023}, its applicability to astrophysically realistic environments and particularly the high-masses of PTA detectable systems \citep[e.g.][]{Bortolas+2021, Siwek+2024} remains uncertain.  Nonetheless, the presence of gas around the MBHs can give rise to several electromagnetic signatures, many producing variations related to the binary's orbital period \citep[e.g.][]{Farris+2014, DOrazio+2015, Munoz+Lai-2016, DOrazio+2018}, these signatures could permit electromagnetic counterparts of any GW-detected system to be identified \citep{2006MmSAI..77..733K, 2013CQGra..30v4013B, DeRosa+2019, Bogdanovic+2022, D'Orazio+Charisi-2023}.

Once the binary has hardened to separations of $\lesssim 10^{-3}$~pc (for the most massive systems), the dominant mechanism influencing its dynamics is the emission of GWs. This emission not only reduces the binary separation (and thus period), but also circularizes the orbit. At these separations, GWs from MBHBs with masses $\sim 10^7-10^{10}\,M_\odot$ can be detected with ensembles of Galactic millisecond pulsars \citep{1995ApJ...446..543R, 2003ApJ...583..616J}. These pulsars emit beams of steep-spectrum radiation along their magnetic field axes, which appear as radio pulses upon intersecting the pulsar-Earth line-of-sight every rotational period. Their exceptional rotational stability and high-timing accuracy can be leveraged to search for minute GW space-time perturbations \citep{Detweiler-1979, Hellings+Downs-1983, Foster+Backer-1990}. The GWs induce fluctuations in the pulsar-Earth proper separation as they transit, thereby advancing or delaying the pulse times of arrival (TOAs). The integrated effect along the pulse geodesic results in two separate terms of influence: an ``Earth term'' from when the GW wavefront passes the Earth (up to angular modulation factors, this term has a common phase between pulsars), and a ``pulsar term'' from when the wavefront passed the pulsar (this is directional and pulsar-distance dependent, recording the orbital phase of the binary from potentially $\sim 10^3$~years in the past) \citep{Detweiler-1979, Jenet+2006}. By searching for the correlated influence of GWs across an ensemble of pulsars that are vastly separated, pulsar timing arrays (PTAs) seek to discover and characterize GWs at frequencies of $\sim1-100$~nHz \citep{2013CQGra..30v4008M}.

There are three PTA collaborations that have been collecting observations for more than a decade: the European PTA \cite[EPTA,][]{2016MNRAS.458.3341D}, the North American Nanohertz Observatory for Gravitational waves \citep[NANOGrav,][]{2013CQGra..30v4008M}, and the Parkes PTA \citep[PPTA,][]{2013PASA...30...17M}. Along with the recently-established Indian PTA \citep{InPTA-2018}, these four collaborations constitute the International Pulsar Timing Array \citep[IPTA,][]{IPTA_DR1-2016}, within which a coordinated effort to combine data and accelerate discoveries is being made \citep{IPTA_DR1-2016, IPTA_DR2-2019}. These collaborations are pursuing three main GW signal classes: $(i)$ a stochastic GW background created through the superposition of many MBHB signals \citep{Rajagopal+Romani-1995, Phinney-2001, Wyithe+Loeb-2003, Sesana+2008}, or of cosmological origin \citep[e.g.][]{Maggiore-2000, Caprini+Figueroa-2018, Christensen-2019}; $(ii)$ individually resolvable ``continuous wave'' MBHB signals that may resound above the background \citep{Sesana+2009, Rosado+2015, Kelley+2018}; and $(iii)$ GW burst signals that follow the ultimate merger of the binary \citep{Zel'dovich+Polnarev-1974, Christodoulou-1991, Wiseman+Will-1991, Thorne-1992}. While the latter may be the only way for PTAs to observe the final coalescence of MBHB evolution \citep{N11yr_memory, N12p5yr_memory}, we do not consider this signal class here. 

Our main targets in this paper are resolvable GW signals from individual MBHBs. While the stochastic GW background (SGWB) is now apparent from recent results \citep{N12p5yr_gwb, N15yr_gwb, PPTA_gwb-2023, EPTA+InPTA_gwb-2023}, PTAs with additional data are expected to be able to resolve several massive, nearby binaries out of the confusion background \citep{2015MNRAS.451.2417R, 2018MNRAS.477..964K, Babak+Sesana-2012}. However, the sky-localization capability of PTAs for threshold binary detections is expected to be relatively poor ($\sim10^2-10^3$~deg$^2$) \citep{Sesana+Vecchio-2010, Taylor+2016, Goldstein+2018, 2024arXiv240604409P}, as is distance determination, rendering host-galaxy identification highly challenging \citep{Goldstein_2019, 2024arXiv240604409P}. Our goal in this paper is to use cosmological hydrodynamical simulations combined with small-scale post-processing of binary dynamics and mock PTA configurations to forecast the demographics of galaxies that could host GW-detectable MBHBs. By doing so, we aim to rank galaxies and thereby drastically reduce the number of plausible hosts that may lie within a GW localization volume from PTA binary detection. Statistical host identification will then trigger dedicated EM follow-up on the smaller subset of likely hosts to search for other morphological, spectroscopic, or photometric signatures of binarity. 

Several previous works have studied different aspects of this problem by conditioning predictions on real galaxy catalogs and simulations. \citet{2014ApJ...784...60S} compiled a $90\%$-complete catalog of galaxies out to $\sim200$~Mpc, injected putative binaries with properties based on galaxy-BH scaling relationships, then ranked potential hotspots based on the binary number density and directional GW power. \citet{2014MNRAS.439.3986R} built on this approach by adapting dark-matter halo clustering and merger metrics developed from the Millennium simulation to a value-added galaxy catalog based on SDSS DR7, of which the latter is limited in completeness and sky coverage. \citet{2017NatAs...1..886M} extended these approaches by populating massive galaxies with $D\lesssim200$~Mpc from the 2MASS survey with putative binaries to assess likely hosts for near-future GW detections, and make empirical predictions for the level of statistical SWGB anisotropy from the local MBHB population. \cite{2021ApJ...914..121A} refined this approach by constructing a 2MASS-based catalog of $\sim44,000$ galaxies out to $D\sim500$~Mpc, and finding that hypothetical binaries within $216$ of those galaxies (mostly elliptical/early-type) lie within the GW-sensitivity volume of the NANOGrav $11$-year Dataset. \citet{2024arXiv240604409P} injected putative MBHBs into a subset of galaxies in different sky regions and at varying distances within this catalog, simulating the entire Bayesian detection, localization, and host-identification process, and applying subsequent cuts based on binary mass and distance consistency between the GW posterior and galaxy catalog values.

From a simulation-driven perspective, \citet{Rosado_2015} generated many realizations of MBHB populations using semi-analytic modeling, and assessed the probability of binary detection as a function of time and PTA configuration. They found that the SGWB was likely to be discovered prior to any individual binaries, but that the detection-probability weighted distributions of these individually-detectable systems favored $\mathcal{M}\sim10^9-10^{10}M_\odot$, $z\sim0.5$, and $f\lesssim10$~nHz. \citet{2019MNRAS.485..248G} used a similar simulation scheme as \citet{Rosado_2015}, and developed a statistical host-ranking metric based on null-stream GW detection statistics and galaxy-BH relationships to reduce the number of plausible hosts by several orders of magnitude. \citet{Kelley_2017a} was the first PTA-study to employ the Illustris cosmological hydrodynamical simulation in which MBHs and galaxies co-evolve.  Those results were used to predict properties of the GWB \citep{Kelley_2017b}, and forecast the detection prospects and properties of individual MBHBs \citep{Kelley_2018} in which they found only a weak dependence on astrophysical conditions such as binary eccentricity and environmental couplings. As expected, the PTA noise properties were found to have a much larger impact.  Especially in the case of substantial pulsar red-noise, these results suggest that single binaries could be detected much sooner than previously expected (\citealt{Kelley_2018}; c.f.~\citealt{Rosado_2015}). 

More recently, \cite{DeGraf_2021} used the Illustris simulation to probe whether the morphologies of recent mergers could indicate the presence of MBHBs, finding that such evidence may persist for $\sim500$~Myrs in PTA-type systems (defined as $\mathcal{M}>10^8 M_\odot$). However, the lag time between galaxy merger and subsequent PTA-detection is such that EM follow-up is unlikely to detect these morphological disturbances. Additionally, they find that the specific star formation rate is enhanced in low-mass mergers, but mostly unaffected in high-mass mergers. However, these results are based on fixed post-merger lag times of $0$, $0.5$, and $1$~Gyr, rather than taking into account distributions of coalescence timescales based on specific galaxy-BH interactions.

While galaxy morphologies may be the most robust indicators of recent galaxy mergers, they are only feasible for a relatively small number of nearby galaxies.  Due to the large distances and thus numbers of possible host galaxies within typical PTA sensitive volumes, it will be crucial to leverage deep, all-sky photometric surveys such as the upcoming Vera Rubin Observatory LSST \citep{Ivezic+2019_lsst}.  The bulk of the billions of galaxies cataloged by LSST will only have point-like measurements of colors and brightness.  Making selections in the standard, galaxy color-magnitude space \footnote{The galaxy color-magnitude diagram is typically presented as r-band absolute magnitude (a proxy for total mass) plotted against g-band minus r-band color (`g-r`, a proxy for typical stellar ages).} \citep[e.g.][]{Bell+2004} is thus the most valuable initial cut on possible host galaxies.

This paper is laid out as follows. In Section~\ref{sec:binary_pops} we outline the selection of MBHBs and host galaxies from the Illustris simulation, as well the details of how we applied sub-grid binary dynamical evolution after the black holes were manually merged by the simulation upon reaching a gravitational softening separation ($\sim 1$~kpc). In Section~\ref{sec:pta_detect} we describe the statistical framework for assessing the PTA detectability of each binary's GW signal, and in Section~\ref{sec:results} we characterize the demographics of galaxy observables for those galaxies that would be statistically-likely hosts for near-future PTA binary detections. We conclude and discuss future prospects in Section~\ref{sec:conclusions}.

\section{Simulated binary population} \label{sec:binary_pops}

\subsection{Merging supermassive black holes within Illustris}

The galaxies and MBH binaries used in our analysis are derived from the Illustris cosmological simulations\footnote{Data for the Illustris simulations are available online at \href{http://www.illustris-project.org/data/}{http://www.illustris-project.org/data/} \citep{Nelson+2015}, and for the newer IllustrisTNG simulations at \href{https://www.tng-project.org/}{https://www.tng-project.org/} \citep{2019ComAC...6....2N}.} \citep{2014Natur.509..177V, 2014MNRAS.444.1518V, 2014MNRAS.445..175G, 2014MNRAS.438.1985T, Nelson+2015}, which are volumes of $(106.5 \, \mathrm{Mpc})^3$ evolved from redshift $z=127$ until $z=0$.  We use the highest resolution ``Illustris-1'' volume, which includes roughly six billion each of gas cells and dark matter particles with mass resolutions of $1.3\times 10^{6} \msol$ and $6.3\times 10^{6} \msol$, and ``softening lengths'' of $710 \, \pc$ and $1400 \, \pc$ respectively.  By redshift zero, the simulation volume contains $\approx 6\times 10^{8}$ stars and over $30,000$ MBHs.

Over the course of the simulation, MBHs are seeded with an initial mass of $\approx 10^5 \msol$ into halos having a mass above $7\times 10^{10} \msol$, and are then allowed to grow via accretion and mergers with other MBHs \citep{2015MNRAS.452..575S}. In order to avoid artificial numerical scattering, MBH particle dynamics are not evolved self-consistently, and are instead re-positioned to the centers of their parent halo.  When two black holes come within a softening length of one another they are ``merged'' instantly. However, in our analysis, we consider this to be the point in time when a binary ``forms'' from the two MBHs\footnote{Note that at the time of ``merger'', in Illustris, the MBH particles are typically separated by $\sim$kpc, and thus are not yet gravitational bound as a true binary, but we continue to use the term for simplicity.}, and we subsequently run the detailed binary evolution in post-processing using semi-analytic models (see Sec.~\ref{sec:mbhb_evo}).  The black hole re-positioning occasionally leads to spurious mergers for MBHs near the seed mass, and thus we only include MBHs with $M > 10^6 \msol$ in our analysis (\citealt{Blecha+2016}; see also the discussion in \citealt{Katz+2020}).

The Illustris simulations are known to accurately reproduce the distributions and properties of not only MBHs but also their host galaxies \citep{2015MNRAS.452..575S, 2014Natur.509..177V}.  In particular, Illustris accurately reproduces the range of galaxy types, and their appropriate correlations with brightness, color, star-formation rate, compactness, and environment \citep{2014MNRAS.444.1518V, 2014MNRAS.445..175G, 2015MNRAS.454.1886S}.  Recently, the IllustrisTNG simulation suite was completed \citep{2019ComAC...6....2N, 2018MNRAS.473.4077P, 2018MNRAS.475..676S}, which focuses on updated models for star-formation driven winds and AGN feedback aimed at better reproducing the inter- and circum- galactic media.  The results from the IllustrisTNG simulations are still consistent with the original Illustris models for the characteristic scalar measurements of galaxies that we employ here (such as mass, magnitude, metallicity, etc).  The TNG models increase the MBH `seed' mass by an order of magnitude, drastically changing the distribution of MBH and MBH mergers properties, thus we have continued to use the original Illustris population to maintain continuity with previous GW studies \citep[e.g.][]{Blecha+2016, 2018MNRAS.477..964K, 2019MNRAS.485.1579K}.

\subsection{Evolving supermassive black-hole binaries into PTA sensitivity band} \label{sec:mbhb_evo}

The Illustris simulations merge MBH particles once they are within the gravitational softening length ($\sim$kpc) of each other and do not have the resolution to include details about the dynamics of MBH mergers. In order to fill in the gaps in detail, we post-processed the Illustris MBH mergers with our own binary evolution scheme which includes both environmental factors and, eventually, gravitational waves.  We summarize the key aspects of our binary evolution model here, but we refer the reader to \citet{Kelley_2017a, Kelley+2017b, Kelley+2018} for more details.   This post-processing begins when the MBHs are ``merged'' by Illustris. The evolution considers the three environmental factors: dynamical friction in the post-merger galactic remnant, stellar loss-cone scattering, and circumbinary disk torques, which allow the system to reach separations necessary to evolve due to the emission of GWs \citep{1980Natur.287..307B, Holley_Bockelmann_2015}.  To calculate these environmental `hardening' mechanisms, the density and velocity profiles from each MBHB host galaxy are extracted from the snapshot before ``merging'' in Illustris, and extrapolated to separations below the simulation resolution. Our current analysis treats all binaries as circular, but eccentric binary evolution can also be included in future analyses.

A detailed description of dynamical friction can be found in \citet{2012ApJ...745...83A}, and the specifics of our implementation are found in \citet{Kelley_2017a}. Stellar scattering is implemented following the formalism and fits to numerical scattering experiments from \citet{2007ApJ...660..546S} which assume a spherical and isotropic background of stars.  When considering circumbinary disk torques, we adopt the three-zone thin accretion disk model from \citet{2009ApJ...700.1952H} which is normalized to the accretion rates derived directly from Illustris.   After evolving due to these environmental factors, the binaries may reach a separation that is small enough such that the hardening will be dominated by the emission of GWs. The GW hardening equations may be found in \citet{1964PhRv..136.1224P}.

Binary evolution \textit{en masse} typically takes billions of years, with only a fraction of binary systems coalescing before redshift zero.  Total binary lifetimes depend sensitively on not only on the masses of the two MBHs, but also the structure and composition of the post-merger host galaxy.  In particular, there are two standard bottlenecks for mergers where some MBHBs can `stall'.  The first is at kiloparsec-scale separations, where tidal stripping of the secondary galaxy can substantially slow dynamical friction, and lead to the secondary MBH being deposited in the primary galaxy's outskirts.  This typically occurs in unequal mass-ratio mergers \citep[$q\equiv m_2/m_1 \lesssim 0.1$][]{Kelley_2017a}.  The second is at parsec separations, the so-called `final-parsec problem' \citep{Milosavljevic+Merritt-2003}, where the population of stars able to scatter with (and extract energy from) the binary becomes depleted.  There is now a general consensus that most galaxies are able to sufficiently replenish this population of stars \citep[e.g.][]{Khan+2011, Vasiliev+2015}, although MBHBs in low total-mass galaxies and those with shallow inner stellar density profiles can still stall \citep{Kelley_2017a}.  Accurate models for binary evolution thus encode important correlations between MBHs/galaxies and which binaries are able to reach the small separations corresponding to detectable GW emission.

Our evolution models do not account for `triple' MBH interactions \citep{Blaes+2002, Hoffman+Loeb-2007}, although these are expected to be subdominant for MBHB mergers overall \citep{Sayeb+2024}.  Additionally, while we do account for circumbinary disk torques, we do not include MBH growth during inspiral \citep{Siwek+2023}; this should not significantly effect binary lifetimes in the PTA regime \citep{Bortolas+2021, Siwek+2024}.

\subsection{Tracking host galaxies over cosmic time within Illustris}

After the two MBH particles are `merged' in Illustris, we identify the host-galaxy in the subsequent simulation snapshot, and the post-processing binary evolution model (described above) yields the age of the Universe at which each binary reaches the PTA sensitivity band.  We then track the evolution of the galaxy using merger trees \citep[`sublink';][]{Rodriguez-Gomez+2015}, until we reach the snapshot immediately following the PTA-detectable epoch. The sublink merger tree for the main descendant branch (sublink\textunderscore mdb) includes arrays of all the snapshot numbers of the descendants of this system, carrying with each of them the ID number at that snapshot.   %\autoref{fig:MergerTreeIllustration} gives a visual depiction of this idea.  
The index of the correct system in the array of snapshots is the system's subhalo ID for that snapshot. There are four systems where the merger tree skips over the snapshot where the desired system should become PTA detectable. This is a workaround in Illustris to address situations in which one subhalo is passing through another at the snapshot of interest \citep{Rodriguez-Gomez+2015}. These are removed from the set of systems we study, but should have a negligible impact on our analysis.

The properties of the host galaxy that we examine are the redshift, galaxy mass, stellar metallicity, specific star formation rate, and color--magnitude relationship. Additionally, we study the chirp mass and frequency of the binary to compare with the other properties of the host galaxy. In order to understand the properties of the host galaxies of systems that are the most likely to be detected, the property histograms must be weighted by some metric to indicate their likelihood of being detected. We choose the detection probability \citep[DP;][]{Rosado_2015} to weight our system, but we compare the DP weighting to weighting by the strain, signal-to-noise, and simply a frequency cutoff for MBH systems which assumes that all systems in the PTA band are detectable.

\begin{figure*}
    \centering
    \includegraphics[width=\textwidth]{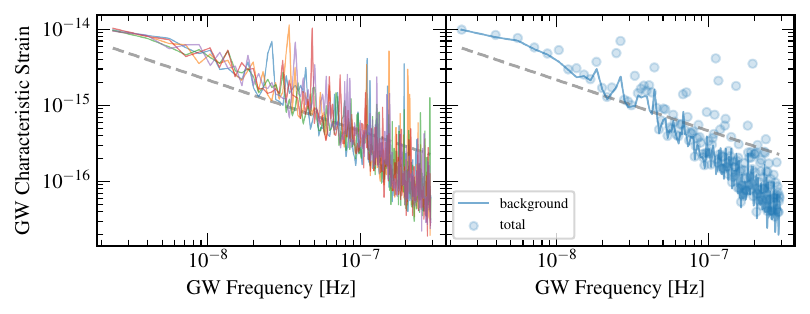}
    \caption{Gravitational-wave characteristic strain spectra from Monte Carlo realizations of MBHB populations resampled from Illustris. Five realizations are shown in the left panel, with each having a different color. The dashed gray line is the expected power-law GWB spectrum assuming binary evolution solely by GW emission, such that $h_c\propto f^{-2/3}$. In the right panel only a single realization is shown, but with the strain spectrum dissected to show the total spectrum as well as the remnant after removing the loudest binary from each frequency bin.}
    \label{fig:Realizations}
\end{figure*}

\section{PTA Detection Probability} \label{sec:pta_detect}

Having tracked the formation of MBHBs in Illustris, then post-processed their dynamical evolution through several stages of environmental interaction until the observed GW frequency falls within the frequency regime of PTA sensitivity, we are left with $N_b=3947$ binaries and host galaxies. We now describe how Monte Carlo realizations of MBHB populations are synthesized from this sample, for which we mostly follow the scheme outlined in \citet{Kelley_2017b}. We then discuss how we assess the probability of detecting GWs from individual binary systems in PTAs.

\subsection{Generating MBHB Populations}

To synthesize Monte Carlo realizations of GW-emitting MBHB populations, we consider the build-up of GW signals from individual systems with chirp mass $\mathcal{M}$ at different observer-frame frequencies $f$, such that the GW characteristic strain can be calculated as,
\begin{align} \label{eq:strain}
    h_c^2(f) &= \int dz d\mathcal{M} \left[ \frac{d^2 N}{dzd\mathcal{M}d\ln f_r} h_s^2(f_r) \right]_{f_r = f(1+z)} \nonumber\\
    &= \int dz d\mathcal{M}\frac{d^2 n_c}{dzd\mathcal{M}} dV_c \left[ \frac{f_r}{df_r}h_s^2(f_r) \right]_{f_r = f(1+z)},
\end{align}
for a number of sources $N$, or a comoving number density $n_c$ in radial comoving volume $V_c$, defined by \citep{Hogg-1999}
\begin{equation}
    dV_c(z) = 4\pi (1+z)^2\frac{c}{H(z)}d^2_c(z)dz,
\end{equation}
where $H(z)$ is the redshift-dependent Hubble factor, $d_c$ is the radial comoving distance, and $f_r$ is the rest-frame GW frequency at redshift $z$.

The discretization of \autoref{eq:strain} over our finite number of sources from the Illustris simulation turns the integral over number density into a sum over sources within the Illustris volume ($V_\mathrm{ill}=(106.5$~$\mathrm{Mpc})^3$):
\begin{equation}
    \int dz d\mathcal{M}\frac{d^2 n_c}{dzd\mathcal{M}} dV_c \rightarrow \sum_{ij} \frac{\Delta V_{ij}}{V_\mathrm{ill}},
\end{equation}
which corresponds to a sum over all binaries $i$ at all simulation time-steps $j$. The volume scaling factor $\Lambda_{ij}\equiv \Delta V_{ij}/V_\mathrm{ill}$ is the number of \textit{astrophysical} MBHBs in the observer's past light cone that is represented by each \textit{simulated} Illustris binary. This depends on the integration step-size $\Delta z_{ij}$,
\begin{equation} \label{eq:lambda}
    \Lambda_{ij} = \frac{1}{V_\mathrm{ill}}\frac{dV_c(z_{ij})}{dz_{ij}}\Delta z_{ij}.
\end{equation}

Our subsequent scheme now differs slightly from \citet{Kelley_2017b}, which associates each Illustris MBHB with a weight drawn from a Poisson distribution $\mathcal{P}(\Lambda_{ij})$, then sums the squared characteristic strain to produce a particular realization of a GWB. Instead, we construct a kernel density estimator\footnote{Using the \texttt{kalepy} package \citep{2021JOSS....6.2784K}.} (KDE) with $\Lambda_{ij}$ as the weight of each simulated binary. This KDE is calculated across $9$ properties of the $N_b$ systems, including those of the MBHB: total mass $M$, mass ratio, $q$, redshift $z$, GW frequency $f$, and those of the host galaxy: $g-r$ color, absolute magnitude ($M_r$), stellar metallicity ($Z$), specific star-formation rate (sSFR), and stellar mass ($M_\star$).  Each of these quantities are then resampled from the KDE using reflective boundary conditions, either determined by physical reasons (e.g. mass ratio $0 < q \leq 1$), or limited to the parameter bounds from Illustris (e.g. restricting total masses to be between the minimum and maximum resolved in the simulation volume).
Thus our distributions of properties are conservative, as they don't extrapolate beyond the limitations of the simulation.

Having constructed our KDE across the $9$-dimensional space of system properties, we re-sample $10^3$ realizations. We draw $\sim2.5\times10^6$ systems in each realization, corresponding to the sum of the $\Lambda_{ij}$ factors, i.e., the total number of astrophysical binaries expected in the observer's past light cone. Using these random draws of system properties, we calculated the characteristic strain for each binary, and the GW background spectrum of each realization. In the left panel of \autoref{fig:Realizations} we show the characteristic strain spectra for five of our $10^3$ realizations, contrasted with the idealized power-law behavior of $\propto f^{-2/3}$ \citep{Phinney-2001}. As expected, GWB spectra from finite-number Monte Carlo realizations depart from the expected behavior at higher frequencies due to low occupancy factors \citep{2008MNRAS.390..192S}. On the right panel of \autoref{fig:Realizations} we show just one realization, where we now exclude the loudest source in each frequency bin as a candidate individually-resolvable MBHB. The remnant GWB signal from all other binaries is treated as a self-noise term in the determination of the loudest source's GW-detection probability. We now discuss how this detection probability is calculated.

\subsection{PTA Detection Statistics}

Our scheme for assessing the PTA detection of GW signals from individual MBHB systems uses the $\mathcal{F}_e$ statistic \citep{2012ApJ...756..175E}. This statistic is based on the assumption that typical binary orbital evolution timescales are shorter than pulsar-Earth light-travel times, such that the frequencies of pulsar-terms will all be different and will lie below that of the common Earth-term frequency. Hence only the Earth-term waveform is used as a matched-filtering template when constructing the PTA likelihood. By forming a ratio of the likelihood in the presence of a signal to one in the absence of a signal, then analytically maximizing this over terms involving the initial GW phase $\Phi_0$, GW polarization angle $\psi$, binary inclination angle, $\iota$, and GW amplitude $A$, we can form a statistic that is only a function of GW sky-location $\{\theta,\phi\}$ and frequency $f$. In the absence of a signal, this statistic is distributed as \citep{2012ApJ...756..175E}
\begin{equation}
    p_0(\mathcal{F}_e)=\mathcal{F}_e e^{-\mathcal{F}_e},
\end{equation}
such that the false alarm probability (FAP) $\alpha_0$ of a signal is given by,
\begin{equation} \label{eq:fap}
    \alpha_0 = 1 - [1 - [ 1+ \bar{\mathcal{F}_e}] e^{-\bar{\mathcal{F}_e}}]^N,
\end{equation}
where $\bar{\mathcal{F}_e}$ is the threshold value of the statistic corresponding to the chosen FAP $\alpha_0$, and $N$ accounts for a number of independent trials of waveform templates that must be performed in order to determine $\{\theta,\phi,f\}$. Following \citet{2014MNRAS.439.3986R}, we set $N=10^4$. We may then numerically solve \autoref{eq:fap} to find the threshold value of $\bar{\mathcal{F}_e}$ for our chosen value of $\alpha_0=0.001$, which was previously adopted by \citet{2014MNRAS.439.3986R} and corresponds to $\sim3$~$\sigma$ significance. While our results would vary quantitatively with different FAP, we expect that our conclusions would not qualitatively change. Detectable gravitational-wave signals from SMBHBs favor massive binaries, high frequencies, and small distances. However, these signal selection effects are convolved with the underlying rate distribution of massive black holes in the Universe. Lowering the FAP would likely select for stronger signals and more extreme system properties, but our choice of FAP here is a reasonable one to avoid spurious noise sources.

In the presence of a signal, the $\mathcal{F}_e$ statistic is distributed as \citep{2012ApJ...756..175E}
\begin{equation} \label{eq:p1}
    p_1(\mathcal{F}_e,\rho) = \sqrt{\frac{2\mathcal{F}_e}{\rho^2}} I_1\left(\rho\sqrt{2\mathcal{F}_e}\right)e^{-\mathcal{F}_e - \frac{1}{2}\rho^2},
\end{equation}
where $I_1(\cdot)$ is a modified Bessel function of the first kind of order $1$, and $\rho$ is the binary's signal-to-noise ratio (S/N). \autoref{eq:p1} can be numerically integrated to compute the detection probability (DP) at a given fixed FAP of a binary, $i$, with a certain SNR:
\begin{equation} \label{eq:dp}
    \gamma_i = \int_{\bar{\mathcal{F}_e}}^{\infty}p_1(\mathcal{F}_e,\rho_i)d\mathcal{F}_e.
\end{equation}
Practically, the function $I_1(\cdot)$ can suffer from numerical overflow when the argument is large. Therefore we use an approximate large-argument analytic expansion whenever numerical overflow occurs. We use this to numerically evaluate \autoref{eq:dp} over a fine grid in S/N, $\rho$, from which we construct an interpolant to map from binary S/N$_i$ to DP$_i$, i.e., $\rho_i\rightarrow\gamma_i$.

The total PTA S/N of a binary adds in quadrature across all pulsars, such that
\begin{equation}
    \rho_i = \left[ \sum_{j=1}^{M} \rho_{i,j}^2 \right]^{1/2}.
\end{equation}
where $\rho_{i,j}$ is the S/N of a binary, $i$, in pulsar, $j$. Dropping the subscript on the $i$-th binary's properties for ease of reading, this is given by
\begin{align} \label{eq:snr}
    &\rho^2_{j} = \frac{A^2}{S_j 8\pi^3 f^3} \{ a^2 [F_j^+]^2 [\Phi_T (1+2\sin^2(\Phi_0)) \nonumber\\
    & + \cos(\Phi_T) (-\sin(\Phi_T) + 4\sin(\Phi_0) ) - 4\sin(\Phi_0)] \nonumber\\
    & + b^2 [F_j^\times]^2 [ \Phi_T (1+2\cos^2(\Phi_0)) + \sin(\Phi_T) ( \cos(\Phi_T)-4\cos(\Phi_0) ) ] \nonumber\\
    & -2ab F_j^+ F_j^\times [ 2\Phi_T\sin(\Phi_0)\cos(\Phi_0) + \sin(\Phi_T)(\sin(\Phi_T)-2\sin(\Phi_0)) \nonumber\\
    & + 2\cos(\Phi_T)\cos(\Phi_0)-2\cos(\Phi_0) ] \},
\end{align}
where $\Phi_T = 2\pi fT$ for a timespan of pulsar observations, $T$; $F_j^{+,\times}$ are the GW polarization response functions for each pulsar, which are a function of the binary's and pulsar's sky location; $a$ and $b$ are functions of the binary orbital inclination-- see \citet{2014MNRAS.439.3986R} for full details and definitions. We distribute all binary positions to be statistically isotropic on the sky, which is well justified given that the influence of galaxy clustering on small scales is likely to be a minor consideration for PTA GW searches \citep{2024arXiv240616031A}. Finally, we calculate $S_j$, the noise power-spectral density of the timing residuals in each pulsar, given by 
\begin{align}
    S_j &= S_{j,\mathrm{white}} + S_{j,\mathrm{red}} + S_\mathrm{GWB} \nonumber\\
    &= 2\sigma_j^2\Delta t + \frac{A_{j,\mathrm{red}}^2}{12\pi^2 f^3}\left( \frac{f}{f_\mathrm{yr}} \right)^{-\gamma_{j,\mathrm{red}}} + \frac{h_{c,\mathrm{rest}}^2}{12\pi^2 f^3},
\end{align}
where $\sigma_j$ is the white-noise RMS in each pulsar, $\Delta t$ is the typical pulsar timing cadence (which we take to be $2$~weeks), $A_{j,\mathrm{red}}$ is the dimensionless intrinsic red noise amplitude in each pulsar, $\gamma_{j,\mathrm{red}}$ is the intrinsic red noise power-spectral density exponent, and $h_{c,\mathrm{rest}}$ is the remainder of the GW characteristic strain signal in each frequency bin upon removing the loudest binary as a candidate resolvable GW signal. The self-noise of the unresolvable population forming a stochastic GWB is thus properly accounted for in our single-source detection statistics.

\subsection{A Near-future PTA}

\begin{figure}
	\includegraphics[width=\columnwidth]{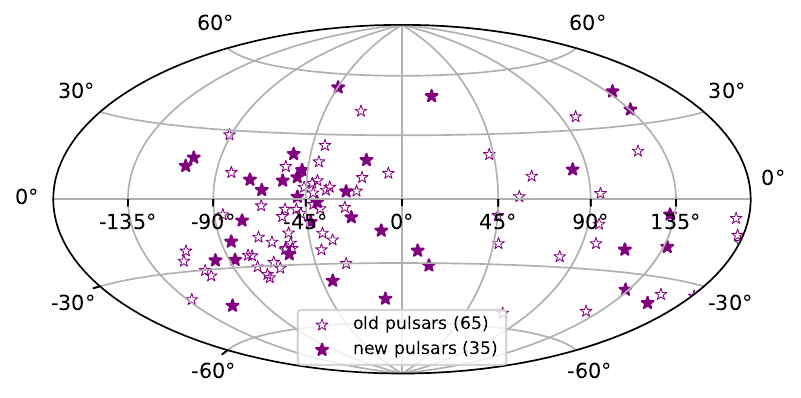}
    \caption{Sky positions of the $100$ pulsars in our near-future PTA. Filled stars indicate the positions of the $65$ pulsars from IPTA Data Release $2$ \citep{IPTA_2nd}. Open stars indicate the positions of $35$ new pulsars that were re-sampled from a KDE fit to the distribution of the original $65$. Our near-future PTA thus follows the distribution of Galactic millisecond pulsars and recent survey areas.}
    \label{fig:Pulsars}
\end{figure}

The final choices needed to compute the DP of a particular binary are the PTA specifications: namely the PTA geometry (number of pulsars and their positions) and the pulsar noise characteristics. We construct a near-future PTA based on the locations and noise properties of $65$ pulsars in the IPTA's $2$nd data release \citep{IPTA_2nd}. We fit a KDE to these locations and re-sample the positions of $35$ new pulsars in order to reflect the distribution of currently timed millisecond pulsars, which itself is a convolution of the Galactic millisecond pulsar distribution and recent survey areas. We thus have a total of $100$~pulsars, which we take to be representative of a near-future PTA that would be operating in the middle to end of this decade. \autoref{fig:Pulsars} illustrates the sky coverage of our PTA. 

Pulsar noise characteristics are selected to be currently realistic, albeit conservative as timing precision will improve in real datasets in the future. We adopt values from scenario-$(c)$ in \citet{Kelley_2018}, which for all pulsars has $\sigma = 0.3\mu\mathrm{s}$, $A_\mathrm{red} = 10^{-15}$, and $\gamma_\mathrm{red}=4.5$. Combined, this corresponds to a total strain noise (i.e., not due to GWs) of $h_{c}^{N}= 6.7 \times 10^{-15}$ at $f=0.1$~yr$^{-1}$. Obviously in real PTAs the noise characteristics are highly heterogeneous, with some pulsars having much better timing precision or lower-frequency sensitivity than others. Likewise, pulsar timing precision improves over time as new radio-timing hardware is brought online or as the duration of timing observations increases. Thus as a first test of our framework to weight MBHB distributions by their detection probability, we also assess the impact of PTA noise characteristics. 

\begin{figure}
	\includegraphics[width=\columnwidth]{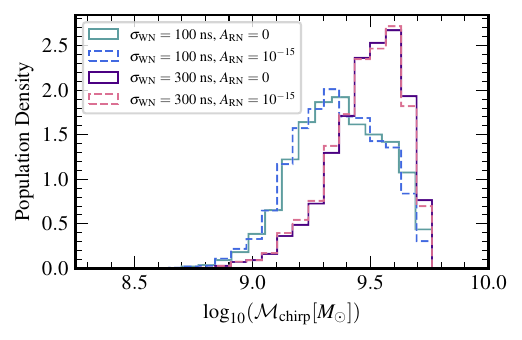}
    \caption{Detection-probability--weighted chirp mass distributions for MBHBs are compared for varying PTA noise characteristics. As expected, higher noise across all frequencies leads to the distribution of detectable binary systems shifting to larger chirp masses which produce higher-amplitude signals. We adopt the $\sigma_\mathrm{WN}=300$~ns, $A_\mathrm{RN}=10^{-15}$ noise characteristics for the rest of this paper.}
    \label{fig:ChirpNoise}
\end{figure}

\autoref{fig:ChirpNoise} shows DP-weighted distributions of binary chirp mass, corresponding to an average over our $10^3$ population realizations. Within each realization, the binary with the largest strain in each frequency resolution bin is selected, and its DP averaged over $10^3$ random draws of possible sky locations, initial phases, and orbital inclinations. This averaged DP is used as a weighting in our distributions. As expected, these favor high chirp-mass values, yet not the highest, since those systems at $\gtrsim 10^{10}M_\odot$ are exceedingly rare and evolve rapidly through the PTA band. We vary the pulsar noise characteristics across $4$ combinations of $\sigma=[0.1,0.3]~\mu\mathrm{s}$ and $A_\mathrm{red}=[0,10^{-15}]$. Surprisingly, we find that the white noise level has the largest overall impact on these distributions, where increased white noise pushes the most likely binary chirp masses to be higher, where they produce stronger, more detectable signals against the raised noise. White noise is broadband, whereas red noise is (by definition) lower frequency. Hence this also indicates that most detectable MBHBs are not found at the very lowest frequencies, but at slightly higher frequencies in the PTA band where the red noise and GWB do not dominate sensitivity.

\section{Results} \label{sec:results}

\begin{table}
  \centering
  \renewcommand{\arraystretch}{1}
  \setlength{\tabcolsep}{4pt}
  \renewcommand{\arraystretch}{1.5}
  \begin{tabular}{c|c}
	\hline\hline
	{\bf Property}                                     &   {\bf Value}                        \\ \hline 
	$g-r$ 	                                           &	$  0.72\substack{+0.11 \\ -0.3 }$ \\
	{Absolute Magnitude, $M_{r}$} 		               & 	$-25.14\substack{+2.64 \\ -1.39}$ \\
	{Stellar Metallicity [$Z_{\odot}$]}	               &	$  2.18\substack{+0.61 \\ -0.41}$ \\
	{sSFR [$10^{-10}\,\mathrm{yr}^{-1}$]}	           &	$  1.06\substack{+1.97 \\ -0.86}$ \\ \hline
	{$\mathcal{M}_\mathrm{chirp} [10^9\, M_{\odot}]$}  &	$  3.55\substack{+1.53 \\ -1.71}$  \\ 
	$z$ 		                                       & 	$  0.86\substack{+0.98 \\ -0.62}$  \\
	$q$	                                               &	$  0.53\substack{+0.37 \\ -0.32}$  \\ \hline\hline
\end{tabular}
\caption{\label{tab:params} Median and $90\%$ range of the observable properties of host galaxies and constituent MBHBs for GW-detectable systems. Values are derived from our GW detection-probability--weighted distributions.}
\end{table} 

In this section we compare the population of host galaxies weighted by the GW detection-probability of their constituent MBHBs, against the population of all binary host galaxies. We remark on how different galaxy observables for hosts of detectable binaries are different from the population of all binary hosts, as well as all galaxies in general. \autoref{tab:params} summarizes our key findings for the median and $90\%$ range of GW detection-weighted population distributions for host galaxies and constituent MBHBs.

\begin{figure*}
    \centering
    \includegraphics[width=\textwidth]{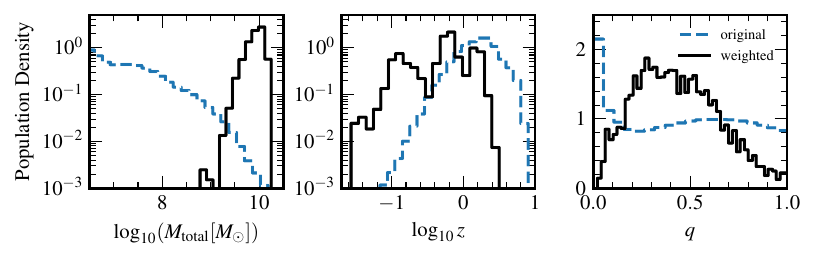}
    \caption{Distribution of MBHB properties for all binaries in our Illustris selection of systems emitting GWs in the PTA band. The blue dashed line corresponds to the unweighted population properties, while solid black lines correspond to weighting systems by their GW detection probability. The \textit{left} panel shows total binary mass, the \textit{center} panel shows the redshift, and the \textit{right} panel shows binary mass ratio, $M_2/M_1$. Note the different vertical scales in each panel for ease of viewing.} 
    \label{fig:MBHBprops}
\end{figure*}

\begin{figure}
    \centering
    \includegraphics[width=\columnwidth]{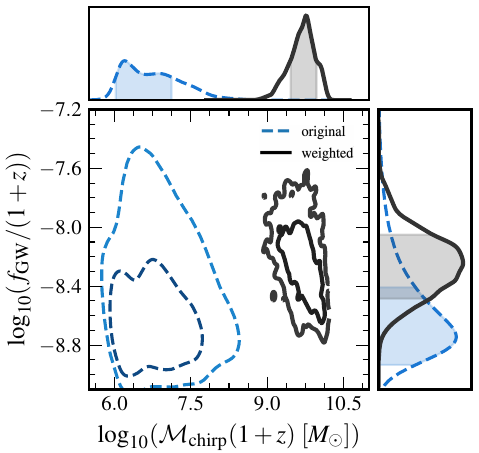}
    \caption{Distributions of binary GW observable properties, specifically the observer-frame chirp mass and the observed GW frequency. As in \autoref{fig:MBHBprops}, the blue dashed lines indicate the distribution for all binaries while the black solid lines correspond to the detection-weighed distribution. Contours enclose $68\%$ and $95\%$ of probability, respectively.}
    \label{fig:GWobs}
\end{figure}

\begin{figure*}
    \centering
    \includegraphics[width=\textwidth]{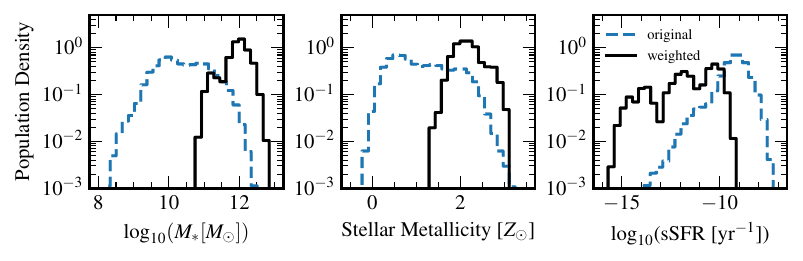}
    \caption{Distribution of host-galaxy properties for all binaries in our Illustris selection of systems emitting GWs in the PTA band. The blue dashed line corresponds to the unweighted host-galaxy population properties, while solid black lines correspond to weighting the host-galaxy distributions by the detectability of their constituent MBHB. The \textit{left} panel shows stellar mass, the \textit{center} panel shows stellar metallicity with respect to the Sun, and the \textit{right} panel shows specific star formation rate.}
    \label{fig:obs_props}
\end{figure*}

\subsection{MBHB Properties}

\autoref{fig:MBHBprops} shows the comparison between the distribution of MBHB properties for all binaries emitting GWs in the PTA band to those of GW-detectable systems. In the left panel, the unweighted population is dominated by lower total-masses, but the selection effects of GW detectability skews this distribution toward higher values around $\sim10^{10}M_\odot$. Likewise, with GW signals having higher strain amplitudes for closer systems (in inverse proportion to their distance), the bulk of the distribution of detectable binaries is $z<1$, peaking at $z\sim0.5$ as shown in the center panel.\footnote{Indeed, even at $z\sim1$, ultra-massive binary BHs in massive clusters could be a source of resolvable signals \citep{2023ApJ...944..112W}.} Finally, the right panel shows that the distribution of mass ratios for all binaries is skewed toward asymmetric systems.  In Illustris, the overall distribution of mass-ratios is relatively flat in log-space.  While the binary evolution model does lead to larger fractions of extreme mass-ratio systems stalling and not reaching the PTA frequency regime, a noticeable fraction do still reach these frequencies \citep{Kelley_2017a}.  In contrast, the detection-weighted distribution peaks around $q\sim0.4$.  One can understand this by recalling that $h\propto\mathcal{M}_\mathrm{chirp}^{5/3}\propto q M_\mathrm{total} / (1+q)^2$, such that for fixed total mass the strain amplitude scales approximately linearly with $q$ at small-$q$ values. Hence GW detectability preferentially selects more equal mass binary systems. We finish this exploration of MBHB properties with \autoref{fig:GWobs}, showing contrasting distributions of GW observables in the form of observer-frame chirp mass and observed GW frequency. The distributions are notably separated, demonstrating how the requirements of GW detectability selects for systems in the high observer-frame chirp mass tail, and preferring observed frequencies $\sim5$~nHz. 

\subsection{Host Galaxy Properties}

% \steve{This section could definitely use your insight, Luke.}
In \autoref{fig:obs_props} we show the analog of \autoref{fig:MBHBprops} for several properties of host galaxies. In the left panel we see that the peak of the stellar mass distribution shifts higher by almost two orders of magnitude from $\sim10^{10}M_\odot$ to $\sim10^{12}M_\odot$ when weighting host galaxies by GW detection probability. In the center panel we show galaxy stellar metallicity, which is a mass-weighted average metallicity of stars within twice the stellar half-mass radius of each Illustris subhalo. The most detectable MBHBs are found in host galaxies where the stellar metallicity is approximately $2Z_\odot$. This is in notable contrast to the overall distribution of binary hosts, which peaks nearer to $0.6Z_\odot$. Galaxies with higher stellar metallicity tend to be older; metallicity increases with subsequent generations of stars, as additional metals are formed through nuclear burning, and ejected into the environment through supernovae and winds to then form the next generation of stars.

Older galaxies are also more likely to have gone through major mergers and thus are more likely to host a MBHB system that is massive enough to be detectable in GWs by PTAs. The fact that the distribution of stellar metallicity becomes noticeably shifted upon weighting by GW detection probability means that it could be one of the best properties for ranking a galaxy's likelihood of hosting a GW-detectable MBHB \citep{Wu_2020}. Finally, in the right panel of \autoref{fig:obs_props} we show specific star formation rate, measuring the rate of star formation per unit stellar mass. Binary hosts in general peak around $\sim10^{-9} \mathrm{yr}^{-1}$, which drops by approximately an order of magnitude or more for galaxies hosting GW-detectable binaries.

\begin{figure}
    \centering
	\includegraphics[width=\columnwidth]{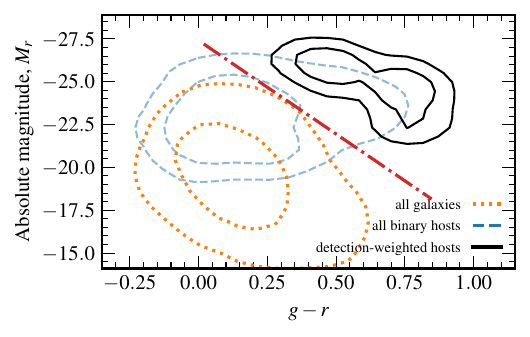}
    \caption{Galaxy properties in the color-magnitude plane for absolute magnitude in $r$-band and $g-r$ color. As in \autoref{fig:MBHBprops} and \autoref{fig:obs_props}, the blue dashed lines indicate the distribution for all binaries while the black solid lines correspond to the detection-weighed distribution. Orange-dotted lines represent the distribution for all galaxies. Contours enclose $68\%$ and $95\%$ of probability, respectively. The galaxies hosting the most GW-detectable MBHBs tend to be redder (high in $g-r$ color) and brighter (more negative in absolute magnitude). The red dash-dot line indicates the decision boundary of a linear support vector machine classifier between the black-solid and orange-dotted distributions.}
    % ((10) * x) - (227/8)
    \label{fig:ColorMag}
\end{figure}

We now turn to galaxy photometric observables, with \autoref{fig:ColorMag} showing the joint distributions of absolute magnitude in the $r$ band, $M_r$ (as an indicator of inherent galaxy brightness) versus $g-r$ galaxy color. As before, blue-dashed and black-solid represent the unweighted distribution for all binary hosts and the distribution weighted by the GW detectability of the constituent MBHB, respectively. The orange-dotted line shows the distribution in the color-magnitude plane for all galaxies, not just binary hosts. Immediately obvious is that galaxies harboring MBHBs that are detectable in GWs by PTAs tend to be brighter and redder than typical binary hosts and typical galaxies overall. This also implies that these galaxies are quenched ellipticals.  We observe that the distribution of binary host properties overlaps with those of all galaxies and the hosts of GW-detectable binaries (as one would expect), but that the compounding requirements of a galaxy hosting a binary that is also detectable in GWs by PTAs leads to virtually negligible overlap in the black-solid and orange-dotted distributions.  It is worth noting that the difference between unweighted binary-hosts and all galaxies is partially due to selection cuts: recall that in Illustris MBH are seeded into halos above a certain mass, which will correlate with brightness.

The red dash-dotted line in \autoref{fig:ColorMag} indicates the decision boundary for a linear support vector machine (SVM) classifier \citep[e.g.][]{SVM-2004} between these latter two distributions, achieving virtually perfect classification (i.e., the area under the receiver operator characteristic curve was $1.0$) on $20\%$ of sampled systems that were held out from training.\footnote{We note that the SVM was trained on equal numbers of sampled systems from each distribution, such that the classification was based on the distribution of properties and not rates.} By contrast, a similar SVM trained to discriminate between binary hosts and all galaxies achieves a true positive rate of $\sim65\%$ with a false positive rate of $0.1\%$, due to the larger (visible) overlap in those distributions. A full systematic investigation of the optimal classification scheme for which galaxy observables can aid in discriminating the hosts of detectable MBHBs is beyond the scope of this paper, but our proof-of-principle analysis here does demonstrate the feasibility of doing so with color--magnitude and other galaxy information.

\section{Conclusions} \label{sec:conclusions}

Individually resolvable single sources producing GW are expected to be detectable by PTAs in the near future. The sky-localization of PTAs is relatively poor and thus, once single sources are detected, it will be unlikely that the true host galaxy will be uniquely pinpointed. Various filtering and selection operations based on consistency between the GW posterior and galaxy catalog values will help reduce the number of plausible hosts \citep{2024arXiv240604409P}, but many possibilities may still remain. It is thus critical to analyze the properties of the host galaxies of the most detectable MBHBs to use as a guide when searching for EM counterparts.

We used the Illustris cosmological simulations coupled with semi-analytic post-processing to identify host galaxies at the time MBHBs become PTA-detectable.  We identified host galaxy properties weighted by the binary detection probability (DP), and compared them to the overall galaxy population.  Using the DP, we found that there was an average of 0.14 systems per realization that have a DP greater than $90\%$. Of the various host galaxy properties, the stellar metallicity has the most contrast between hosts of detectable binaries and that of all galaxies. The original unweighted population peaks at $\sim$0.6 solar metallicities while the DP weighted distribution peaks at twice solar metallicity. The color--magnitude diagram gives an indication of the visual appearance of the most detectable galaxies. They are far more red and far brighter, indicating they are most likely large elliptical galaxies with significantly diminished star formation.

Each of our distributions of host galaxy properties gives us a tool to rank prospective galaxies by their likeliness of harboring a PTA-detected single source; this is very useful information to aid the goal of multi-messenger astronomy of MBHBs. This demographic profile of binary hosts could be combined with other recent work showing that galaxy morphology \citep{2024ApJ...961...34B} and stellar kinematics \citep{2024arXiv240714061B} could signpost the presence of a massive central binary. A natural next step for future work is to investigate the properties of MBHBs that would produce detectable electromagnetic counterpart signatures \citep[e.g.,][in an SKA context]{2023A&A...676A.115P}.  For a review of these counterparts, we point the reader to \citet{D'Orazio+Charisi-2023}.  Some of the most promising signatures are shifted broad-line regions \citep[e.g.][]{Kelley-2021}, accretion variability \citep[e.g.][]{Farris+2014}, Dopper boosting \citep[e.g.][]{D'Orazio+2015}, and self-lensing \citep[e.g.][]{D'Orazio+2018, Kelley+2021}.  However, due to the extreme intrinsic variability of AGN, each of these signatures is subject to substantial contamination from false positives. Side by side with the population that is detectable in GWs by PTAs---as we have investigated in this work---we can thus drastically improve the chances of identifying multi-messenger sources in the next few years.
% examine the typical properties of binaries (and their hosts) that are the most likely to be multi-messenger detectable. 

\section*{Acknowledgments}

We thank the anonymous referees for their comments. We thank Maria Charisi for extensive discussions that improved the quality of this work. KC acknowledges support from the Vanderbilt University Data Science Institute Summer Research Program (DSI-SRP). SRT acknowledges support from NSF AST-2007993, AST-2307719, and an NSF CAREER PHY-2146016, and a Vanderbilt University College of Arts \& Science Dean's Faculty Fellowship. SRT and LK are members of the NANOGrav collaboration, which receives support from NSF Physics Frontiers Center award number 1430284 and 2020265. This work was conducted in part using the resources of the Advanced Computing Center for Research and Education (ACCRE) at Vanderbilt University, Nashville, TN.

%%%%%%%%%%%%%%%%%%%%%%%%%%%%%%%%%%%%%%%%%%%%%%%%%%
%\section*{Data Availability}

%The inclusion of a Data Availability Statement is a requirement for articles published in MNRAS. Data Availability Statements provide a standardised format for readers to understand the availability of data underlying the research results described in the article. The statement may refer to original data generated in the course of the study or to third-party data analysed in the article. The statement should describe and provide means of access, where possible, by linking to the data or providing the required accession numbers for the relevant databases or DOIs.



\bibliographystyle{mnras}
\bibliography{bib} % if your bibtex file is called example.bib

%\begin{thebibliography}{99}
%\end{thebibliography}

%%%%%%%%%%%%%%%%%%%%%%%%%%%%%%%%%%%%%%%%%%%%%%%%%%

%%%%%%%%%%%%%%%%% APPENDICES %%%%%%%%%%%%%%%%%%%%%

% \appendix

% \section{Some extra material}

% If you want to present additional material which would interrupt the flow of the main paper,
% it can be placed in an Appendix which appears after the list of references.

% \citep{Charisi_2016}

%%%%%%%%%%%%%%%%%%%%%%%%%%%%%%%%%%%%%%%%%%%%%%%%%%

% Don't change these lines
%\bsp	% typesetting comment
%\label{lastpage}
\end{document}